\documentclass[usenatbib,fleqn]{mn2e}
\usepackage{amsmath}
\usepackage{psfig}

\newcommand{\apj}{ApJ}
\newcommand{\aj}{AJ}
\newcommand{\mnras}{MNRAS}
\newcommand{\aap}{A\&A}
\newcommand{\apjs}{ApJS}

\newcommand{\nature}{Nat}
\newcommand{\phr}{Phys.Rep.}

\def\per#1{{#1$^{-1}$}}
\def\Msol{{$M_{\odot}$}}
\def\micron{{$\mu \rm{m}$}}
\def\etal{{et~al.}}
\def\gsim{\mathrel{\hbox{\rlap{\lower.55ex \hbox {$\sim$}}\kern-.0em\raise.4ex \hbox{$>$}}}}
\def\lsim{\mathrel{\hbox{\rlap{\lower.55ex \hbox {$\sim$}}\kern-.0em\raise.4ex \hbox{$<$}}}}

\title[Submm properties of GRB host galaxies]
{The submm properties of GRB host galaxies}

\author[N.~R.~Tanvir et al.]
{N.~R.~Tanvir$^1$, 
V.~E.~Barnard$^2$, 
A.~W.~Blain$^3$, 
A.~S.~Fruchter$^4$,
C.~Kouveliotou$^5$,\newauthor 
P.~Natarajan$^{6}$, 
E.~Ramirez-Ruiz$^{7}$,
E.~Rol$^8$, 
I.~A.~Smith$^9$,
R.~P.~J.~Tilanus$^2$,\newauthor 
and R.~A.~M.~J.~Wijers$^{10}$.
\\
\\
$^1$Centre for Astrophysics Research, University of Hertfordshire, College Lane, Hatfield, AL10 9AB, UK\\
$^2$Joint Astronomy Centre, 660 N, A'ohoku Place, Hilo, HI 96720, USA\\
$^3$Dept. of Astronomy, Caltech 105-24, Pasadena, CA 91125, USA\\
$^4$Space Telescope Science Institute, 3700 San Martin Drive, Baltimore, MD21218, USA\\
$^5$NASA Marshall Space Flight Center, SD-50, NSSTC, 320 Sparkman Drive, Huntsville, AL 35805, USA\\
$^6$Dept. of Astronomy, Yale University, New Haven, CT 06250-8181, USA\\
$^7$Institute for Advanced Study, Einstein Drive, Princeton, NJ 08540, USA\\
$^8$Osservatorio Astronomico di Padova, Vicolo dell'Osservatorio 5, 35122, Padova, Italy \\
$^9$Dept. of Physics and Astronomy, Rice University MS 108, Houston, TX 77005-1892, USA\\
$^{10}$Astronomical Insitute `Anton Pannekoek', University of Amsterdam and Center for High-Energy Astrophysics, \\
~Kruislaan 403, 1098 SJ Amsterdam, The Netherlands\\
}

\begin{document}
\maketitle

\begin{abstract}

Long duration gamma-ray bursts (GRBs) accompany the deaths of some
massive stars and hence, since massive stars are short lived, are a
tracer of star formation activity.  Given that GRBs are bright enough
to be seen to very high redshifts, and detected even in dusty
environments, they should therefore provide a powerful probe of the
global star formation history of the universe.  The potential of this
approach can be investigated via submm photometry of GRB host
galaxies.  Submm luminosity also correlates with star formation rate,
so the distribution of host galaxy submm fluxes should allow us to
test the two methods for consistency.  Here, we report new JCMT/SCUBA
850\micron\ measurements for 15 GRB hosts.  Combining these data with
results from previous studies we construct a sample of 21 hosts with
$<1.4$~mJy errors.  We show that the distribution of apparent
850\micron\ flux densities of this sample is reasonably consistent
with model predictions, but there is tentative evidence of a dearth of
submm bright ($>4$~mJy) galaxies.  Furthermore, the optical/infrared
properties of the submm brightest GRB hosts are not typical of the
galaxy population selected in submm surveys, although the sample size
is still small.  Possible selection effects and physical mechanisms
which may explain these discrepancies are discussed.

\end{abstract}


\section{Introduction}
\label{sec:intro} 

\subsection {GRBs and star formation}

The spectroscopic detection of an energetic supernova (SN2003dh)
concurrent with GRB~030329 (Hjorth \etal\ 2003a; Stanek \etal\ 2003)
has firmly established that long-duration ($>2$~s; Kouveliotou \etal\
1993) GRBs accompany the core-collapse of some class of massive stars.

Since massive stars are short-lived, this also confirms that GRBs are
closely associated with star formation activity; a possibility already
discussed by a number of authors (eg. Wijers \etal\ 1998; Totani 1999;
Blain and Natarajan 2000).  The extreme luminosity of the prompt
gamma-ray emission means that GRBs can be detected, if they exist, to
very high redshifts, with minimal extinction by intervening gas or
dust.  This makes them potentially very powerful indicators of star
formation to early times.

To date, spectroscopically-confirmed 
redshifts have only been published
for about three dozen bursts, although
various schemes have been suggested 
to derive 
redshifts empirically from $\gamma$-ray properties;
such as the
lag-luminosity (Norris \etal\ 2000) and variability-luminosity
relations (Reichart \etal\ 2001).  
These studies suggest that the
redshift distribution of GRBs is 
broadly consistent with the emerging
picture of the comoving star formation rate in the universe having
peaked sometime around redshifts 1-4, although uncertainties
ramp up at higher redshifts.  (eg. Ramirez-Ruiz, Trentham and
Blain 2002; 
Lloyd-Ronning \etal\ 2002)

These results are interesting, but premature to 
the extent that we have limited
knowledge of these luminosity correlations, and only rudimentary
understanding of the relationship between GRB rate and star formation
rate (eg. Krumholz \etal\ 1998).  
For instance, there is the possibility that GRB rate and/or
brightness depends also on other factors, such as metallicity,
galactic environment, and certainly on any variations in the
stellar initial mass function
(IMF).  These are difficult factors to disentangle, but
important insights can be gained by comparing GRB rate with the star
formation rate estimated by other means.  
Our program
is aimed at providing a more quantitative comparison of the
star formation rate deduced by GRBs and that obtained from submm luminosity, 
through direct study of
the host galaxies of GRBs.  

Existing methods of mapping the star formation history of the universe
rely on estimating the star formation rates of individual galaxies,
and summing these up in redshift bins with some estimated correction
for galaxies below the detection threshold (eg. Adelberger \& Steidel
2000), or by modelling the redshift distribution so as to fit
integrated backgrounds (eg. Blain \etal 1999a).

If GRBs reliably trace star formation, and we can quantify the
relationship, then ultimately detection of their host galaxies will
not be required for the purposes of mapping global star formation
history.  Nonetheless, GRB hosts also uniquely allow us to study the
star-forming galaxy luminosity function right to the faint end.  An
upshot may be a means to estimate the proportion of the total star
formation which is going on in IR-bright, dusty galaxies,
optically-bright galaxies and in fainter populations which are not
selected in other surveys (eg. Trentham \etal\ 2002).  To date, the
best example of the power of GRB selection to probe the faintest end
of the galaxy luminosity function is GRB~020124, whose host galaxy is
undetected by {\it Hubble Space Telescope (HST)} to $R=29.5$ (Berger \etal\ 2002).  This galaxy
would not have been found in any direct imaging survey to date and yet
observations of the afterglow show that the host has a high HI column
density (making it a damped Ly$\alpha$ absorber) and a redshift of
$z=3.20$ (Hjorth \etal\ 2003b).

\subsection {Mapping star formation in the submm}

The power of submm studies for mapping the star formation history of
the universe is discussed in detail by Blain \etal\ (2002) and Smail
\etal\ (2002).  Briefly, in dusty systems UV radiation, predominantly
from massive stars, is reprocessed by the dust and emitted in the far
IR.  This emission, which itself is unaffected by dust extinction, is
thus proportional to the obscured star formation rate.  At higher
redshifts the peak of emission moves increasingly into the submm, with
the beneficial consequence that at 850\micron\ the apparent luminosity
of a galaxy of given intrinsic bolometric luminosity changes little
from redshift $z\approx0.5$ out to $z\approx 10$.

Submm surveys, combined with constraints from the intensity and
spectrum of IR backgrounds, show that compared to the local universe,
the majority of high redshift star formation appears to be taking
place in dusty systems (Blain \etal\ 1999a), and much (although not the
dominant part) of this in so-called ultraluminous infrared galaxies
(ULIRGs: with IR luminosities $\gsim10^{12}L_{\odot}$).  Low redshift
ULIRGs exist, but have around 1000 times lower comoving space density
(Smail \etal\ 2002).

Potential drawbacks with submm surveys are: (a) 
at the 2~mJy confusion limit of SCUBA
only about 30\% of the total submm background
emission from {\it
COBE}-FIRAS observations is resolved out;
(b) any individual galaxy may suffer some contamination of its
850\micron\ luminosity through heating from an obscured AGN.  
Only about 10\% of
submm galaxies with deep X-ray data appear to show evidence for a hard
X-ray AGN (Almaini \etal\ 2003), but larger samples and deeper
observations will be necessary to confirm this fraction; (c) it is
necessary to assume or constrain the shape of the far-IR spectral
energy distribution (SED) at wavelengths shorter than 850\micron\ 
in order to translate the measured submm flux density into an
accurate luminosity and star formation rate. The luminosity inferred
from a galaxy with a certain SED depends only weakly on redshift;
however, without direct knowledge of the details of the SED this
luminosity is uncertain, leading to ambiguity in the results. A
certain fractional change in the dust temperature leads to a
fractional uncertainty in the inferred luminosity that is greater by
several times (Blain \etal\ 2002; Blain, Barnard and Chapman 2003).  
While results are so far generally
consistent with dust temperatures of order 35\,K (Chapman \etal\
2003a), the extent of the distribution of values is not yet known.

Redshifts for large samples of submm galaxies were for a long time
hard to obtain, because of the lack of both bright optical
counterparts (frequently $I>26$), and the poor 15~arcsec spatial
resolution of SCUBA at 850\micron.  However, Chapman et al.  (2003b;
see also Ivison \etal\ 2002) have demonstrated that about 65\% of
submm galaxies brighter than 5\,mJy are detected in very deep VLA
radio maps, providing accurate subarcsec positions. Redshifts for
around 40--50\% of these radio-detected submm galaxies can be obtained
using Keck LRIS spectra (Chapman \etal\ 2003a). Based on these results
they find a range of redshifts from 0.8 to 2.8 with a median of 2.4,
and conclude it is likely that most submm galaxies lie at redshifts
between 2 and 3.  The radio selection could be biased to lower
redshifts, and to sources containing AGN; however, the reasonable
$\simeq 40$\% completeness of Chapman \etal 's redshift determinations
suggests that these effects are not too significant.

On the other hand, in a recent study of the evolution of the global
stellar mass density $0<z<3$ based on an infrared-selected sample of
galaxies from the Hubble Deep Field North, Dickinson et al. (2003)
report that while the star formation rate essentially tracks that
determined in other wave-bands at low redshifts, the rate at $z > 2$
is significantly different.  These observations appear to be
inconsistent with scenarios in which the bulk of stars in present-day
galactic spheroids formed at $z>>2$, since most of the stars (50 --
75\%) of the present day stellar mass density formed by $z\sim1$ and
in fact by $z = 2.7$ only 3 -- 14\% of today's stars were present.
The issue is clearly not settled.

Current sensitivities and confusion limits imply that substantially
fainter submm samples will not be studied in detail until the ALMA
interferometer is commissioned in 2012. Hence, an alternative method
to probe the star formation in obscured galaxies is definitely
required, and GRB source counts and host galaxy studies may ultimately
be able to tell us both when and where this star formation occured.

\subsection {Pros and cons of GRBs as star formation indicators}

Several characteristics of GRBs lend them to probing high
redshift star formation:
(a) they are bright enough to be seen to $z\sim20$ (eg. Lamb \& Reichart 2000);
(b) they can be detected in $\gamma$-rays 
through large columns of dust and gas; 
(c) they can be 
detected independently of whether a host galaxy can be found.
Furthermore, much information relating to the host, such as
redshift, metallicity, gas column density
and extinction 
can all be obtained indirectly,
from afterglow observations --
either optically or x-ray;
(d) the spectral slopes of both prompt and afterglow emission
compensate to some extent for redshift dimming,
and time-dilation means that observations can be made
earlier in the rest-frame time than would be the case
for lower redshifts; 
(e) being produced by individual
massive stars, they are obviously unaffected by
AGN contamination.

On the downside, GRBs are rare, 
their progenitors are still not fully understood and
they are not very useful for telling us about the 
star formation rate in individual galaxies.
In terms of the current observational state-of-play,
samples of GRBs, particularly those with firm redshifts
are very inhomogeneous -- the result of a wide variety of triggers and
followup campaigns.  
One consequence is that, of the roughly
200 entries to-date in Jochen Greiner's web table\footnote{
http://www.mpe.mpg.de/$\sim$jcg/grbgen.html} of ``well-localised'' bursts,
fewer than 25\% have  optical afterglows identified.
However, only a minority of those
200 were reported within 24 hours, and had error
circles $<10$~arcmin in diameter.
Also, many were not well placed for
optical observation, have only shallow limits, or in some
cases no reported optical followup at all.
Thus, the superficially high rate of ``dark'' GRBs is
misleading.
The proportion of genuinely dark GRBs (a reasonable
working  definition would be $R>23.5$ at 24 hours post-burst; see
also section 2) amongst the
{\it HETE} and {\it BeppoSAX} triggers is probably only 10--30\%.

The wide variation in GRB followup campaigns makes
it hard to quantify selection effects, but selection effects
may well be important for our study.
To be found optically, an afterglow
should not be in too dusty an environment. 
A low redshift illustration of this issue is provided
by Mannucci \etal\ (2003), 
who find an enhanced core-collapse supernova rate in
K-band monitoring of nearby star-forming galaxies, but still
conclude that the large majority of supernovae remain 
undetected due
to very high extinction.
In fact, for GRBs this
conclusion is not as inevitable as it sounds because the initial flux of
high energy photons from GRBs is expected to destroy dust possibly up
to $\sim 100$~pc (eg. Fruchter \etal\ 2001c; Galama and Wijers 2001).  
In some cases this will be enough to create a window
through otherwise obscuring dust clouds.
Of course, if most star formation in the universe is occuring
in the ``obscured mode'', then dust destruction may 
actually be a {\em requirement}
to explain the large number of afterglows detected optically.
Even when optical afterglows are not found to faint limits,
as we shall argue below, it is not necessarily the case
that 
they are heavily enshrouded in dust.
Nonetheless, it would be surprising if some fraction of
GRB afterglows were not missed because they were highly extinguished.

Detected afterglows are unlikely to 
be found in very low-density environments,  even if GRBs
occur there, since
the brightness of the spectral peak reduces with the density
of the ambient medium
(Sari \etal\ 1998).  The high HI column densities
seen towards many bursts 
(eg. Galama \& Wijers 2001; Hjorth \etal\ 2003b)
are consistent with this expectation, although
Reichart \& Price (2002) suggest that for the limited
sample of dark bursts available, the column densities are not
consistent with them being in the nuclear regions
of ULIRGs either.

Observational limitations are also likely to introduce
a bias
against finding
high redshift GRBs.
Although the spectral slope and cosmological time-dilation
work so as to reduce the effect of redshift on the magnitude
of an afterglow at a given observer time after the burst (another kind
of negative $k$-correction), 
GRBs of comparable intrinsic luminosity will 
still appear fainter at higher
redshift, making them and their afterglows harder to
discover.
For example,
Hogg and Fruchter (1999) adopted a $(1+z)^{-1}$ dependence for the
probability that a burst at a given redshift is detected
and an optical afterglow found.
Of course,
 $z\gsim7$ objects will
be essentially invisible in the optical due
to the Ly$\alpha$ break.

Our initial goal is to compare the submm properties of GRB hosts with
model predictions
and hence provide a consistency check on both techniques for 
tracing star formation.
Ultimately we'd
like to understand the quantitative relation between GRB rate and star
formation rate, so that larger samples of GRBs may be used to give a
good description of the global star formation history of the universe.

\section{Previous submm studies of GRB hosts}
\label{sec:previous}

The first submm limits and detections of GRB host galaxies came from
observations aimed at detecting GRB afterglows.  Only
in a couple of cases were such fading
afterglows detected, but GRB~010222 was
found to be a steady source, suggesting the flux was dominated
by emission from a bright host galaxy  (Frail \etal\ 2002).
Useful upper limits were also obtained for a number of other hosts
(Smith \etal\ 1999; Galama \etal\ 1999; Kulkarni \etal\ 1999b; Smith \etal\ 2001).

It is possible to average a number of non-detections providing
systematic uncertainties are small, to find the average flux density
for a whole sample.  
Unfortunately in the case of the
afterglow observations, since there may well be some small afterglow
contribution to each observation,
even if it is not detected significantly in individual cases, 
this averaging 
procedure is best avoided.  
For the data discussed here
the observations were generally made long after the
afterglow should have faded.

An obvious concern, already raised in
section 1.3, is that those GRBs with optically detected
afterglows may be biased against residing in dusty hosts, the very
galaxies which would on average be submm bright.  
In an effort to assess
whether such a bias exists, Barnard \etal\ (2003; hereafter paper 1)
 observed a small sample of ``dark''
GRBs with deep limits on any optical afterglow
but with
good radio and/or X-ray positions.
The expection at the outset was that compared to the
hosts of optically-bright GRBs, these
would be more likely to be dusty, massive star-forming galaxies.
In fact, this sample of  4
produced only 1 individually significant detection, and no
overall excess of 850\micron\ emission over that predicted by 
Ramirez-Ruiz, Trentham \& Blain (2002)  for all hosts.

These results appear to argue against dark bursts being
preferentially found in dusty hosts.
However, this is only a small sample,
and in paper 1 we also remarked that there was some evidence
from the rapid decay of the radio afterglows
that two of the bursts in submm-faint hosts could have been dark
due to intrinsic optical faintness, rather than extinction.
In fact, 
it is becoming clearer that there is a broad range in brightness
of optical afterglows and
a number of detected bursts would have been missed in
most afterglow searches.  For example, GRB~980613, GRB~000630
and GRB~021211 (Hjorth \etal\ 2002; Fynbo \etal\ 2001,
Fox \etal\ 2003) were all around $R\sim23$ at 1 day,
and, although the sample of bursts
studied in paper 1 were fainter than all these, the 
upper limits do not require them to have been much fainter.
Similarly, GRB~020124 had a relatively typical intrinsic magnitude, 
but appeared faint because of its redshift of
$z=3.2$ (Hjorth \etal\ 2003b).
In any event, as we see below, 
even the modest rate of submm detections for dark burst hosts
found in paper 1
appears to be somewhat greater
than the rate which is now found for hosts of 
optically detected bursts.

Recently Berger \etal\ (2003; hereafter Be03) 
published SCUBA
850\micron\ measures for a larger sample of 13 GRB hosts, in 
addition to radio observations for many.  Below we present our
results for another sample of hosts (which partially overlaps
with Be03), and combine and
analyse all the extant data.

\section{New data}
\label{sec:newdata}

We have obtained further submm photometry of the host galaxies 
of optically identified GRBs,
using
the Sub-millimetre Common User Bollometer Array (SCUBA) instrument on
the James Clerk Maxwell Telescope (JCMT).
The targets were chosen to be well placed for observation from Mauna Kea, and
have sub-arcsec positions from their optical afterglows.
All observations were made at least 12 weeks after the burst
(and usually much longer),
and are therefore very unlikely to be contaminated with
afterglow emission.

Observations and reduction were performed as described in paper 1.
The log and results are detailed in table 1; note that some were
reported by Barnard (2002).
Of the galaxies presented here and in paper 1, seven are in common
with Be03.  This is useful to improve the measurement uncertainties,
and also to check for statistical consistency, since the methods of
reduction differ somewhat.
Overall, we are reassured to find no significant
systematic
difference between the two data sets.  The galaxy to galaxy scatter
is, however, a little  larger than expected from the quoted errors.
Specifically, 4 of the 7 disagree by more than $1\sigma$, which
could happen by chance, but it 
leads us
to suspect that the uncertainties found by one or both groups
are marginally underestimated.

\begin{table}

\caption[Photometry observations of the new GRB host sample]{ 850\micron\
photometry observations of a new sample of GRB hosts, chosen on the basis of
good positions easily observable from Mauna Kea.  Note the varying
number of integrations per target -- the quoted flux densities
are error-weighted averages of the measurements for each source.  
The $\tau_{~850~\mu\rm{m}}$ measures the sky opacity
during the observations.
As can be seen, we do not
enforce positivity on the submm fluxes, so as not to bias any
subsequent statistical analysis.  In a few cases the uncertainties are
still large, and the data of limited use, but we report them here for
completeness.  
}
\begin{center}
\begin{tabular}{lcrcr}\hline
GRB & Obs. date  & Int. (s) &   
$\tau_{~850~\mu\rm{m}}$ 
& Flux density (mJy)\\\hline
970228 & 23/09/02 & 2700 & 0.30 & 1.78 $\pm$ 1.32 \\
       & 26/12/02 & 1350 & 0.14 & \\
980326 & 24/03/02 & 1800 & 0.28 & -0.27 $\pm$ 1.18\\
       & 23/09/02 & 2700 & 0.30 & \\
       & 28/12/02 & 1350 & 0.09 & \\
980329 & 31/03/02 & 450  & 0.33 & -1.53 $\pm$ 1.19\\
       & 20/09/02 & 900  & 0.32 & \\
       & 22/09/02 & 1800 & 0.25 & \\
       & 03/10/02 & 1350 & 0.33 & \\
       & 23/12/02 & 1350 & 0.17 & \\
980703 & 22/09/02 & 1350 & 0.22 & -1.36 $\pm$ 1.14\\
       & 23/09/02 & 1350 & 0.28 & \\
990123 & 21/04/02 & 450 & 0.35 & -4.18 $\pm$ 4.55\\
990308 & 30/03/02 & 324 & 0.25 & 0.02 $\pm$ 1.75\\
       & 08/12/02 & 1350 & 0.12 & \\
991208 & 22/03/02 & 3330 & 0.16 & 1.97 $\pm$ 1.22\\
000301C & 21/03/02 & 4050 & 0.20 & -1.81 $\pm$ 1.21\\
000926 & 22/03/02 & 1350 & 0.16 & 1.40 $\pm$ 1.23\\
           & 30/03/02 & 675 & 0.15 & \\
           & 26/04/02 & 2250 & 0.19 & \\
001025A* & 05/10/01 & 2250 & 0.30 & -2.53 $\pm$ 3.04\\
010921 & 26/04/02 & 2502 & 0.20 & 0.46 $\pm$ 1.14\\
       & 23/09/02 & 2700 & 0.30 & \\
       & 10/12/02 &  450 & 0.12 & \\
011211 & 12/03/02 & 1260 & 0.40 & 3.81 $\pm$ 1.87\\
           & 19/03/02 & 2700 & 0.32 & \\
020124 & 02/01/03 & 1350 & 0.26 & 1.20 $\pm$ 2.30\\
020813 & 30/12/02 & 1350 & 0.22 & -1.40 $\pm$ 3.50\\
021004 & 29/12/02 & 2250 & 0.14 & 0.77 $\pm$ 1.25\\
       & 30/12/02 & 1350 & 0.22 & \\
\hline
\end {tabular}
\label{obstable}
\end{center}
\noindent
{\footnotesize * The position of
GRB~001025A was first observed as part
of the dark burst program reported in paper 1, but was subsequently
rejected because of doubts over the validity of the X-ray afterglow
identification.  More recently the original identification has been confirmed
(Watson \etal\ 2002), so we report this submm measurement here.}
\end{table}

In table 2, we combine
these results with the data presented
in paper 1 and Be03.  Where galaxies have been observed
twice, we average using weights derived from the quoted
errors.  In one case, GRB~000911, Smith \etal\ (2001)
find an flux which is inconsistent with Be03, and we
therefore average those results together.
In another case, GRB~990123, the limits from Galama \etal\ (1999)
and Kulkarni \etal\ (1999b)
on afterglow emission are a considerably stronger constraint on any
host contribution than our photometry, so again we adopt an average 
value.  Although strictly
this could introduce 
afterglow contamination
it must be very small.

\begin{table*}
\caption{Compilation of GRB hosts with submm observations
from our program and Berger \etal\ (2003).
The fluxes are weighted means
of the available photometry.  
Most of these galaxies are not detected significantly
in their own right and, as expected, some formally
have negative fluxes, simply due to noise.
Redshifts and magnitudes of host
galaxies have been obtained from the literature, with the latter being 
converted to the Cousins $R$ system where necessary and
corrected for foreground extinction according to the Schlegel \etal\ (1998)
maps.
} 
{\vskip 0.75mm} 
%
%
%

\begin{tabular}{lrrcl}\hline
GRB & Redshift & Host $R_C$  & 850\micron\ flux density  & Notes and references\\
host &          & mag.  &  \hfil(mJy)\hfil  &  \\ \hline
970228 & 0.70 & 24.6 &  0.20 $\pm$ 0.81   & Bloom \etal\ 2001; Fruchter \etal\ 1999a; Galama \etal\ 2000  \\
970508 & 0.84 & 25.1 & -1.57 $\pm$ 1.01   & Metzger \etal\ 1997; Fruchter \etal\ 2000b \\
970828 & 0.96 & 25.1 &  1.26 $\pm$ 2.36   & Dark; Djorgovski \etal\ 2001  \\
971214 & 3.42 & 25.6 &  0.49 $\pm$ 1.11   & Kulkarni \etal\ 1998  \\
980326 & ?     & 27.9  & -0.27 $\pm$ 1.18  & Bloom \etal\ 1998; Fruchter, Vreeswijk \& Nugent 2001  \\
980329 & ?     & 26.3 &  0.71 $\pm$ 0.69   & Jaunsen \etal\ 2003 \\
980613 & 1.10 & 23.9 &  1.75 $\pm$ 0.92   & Djorgovski \etal\ 2003; Hjorth \etal\ 2002 \\
980703 & 0.97 & 22.6 & -1.53 $\pm$ 0.72   & Djorgovski \etal\ 1998; Holland \etal\ 2001 \\
981226 & ?     & 24.8  & -2.79 $\pm$ 1.17  & Dark; Frail \etal\ 1999 \\
990123* & 1.60 & 23.9 & 0.47 $\pm$ 0.60   & Bloom \etal\ 1999; Kulkarni \etal\ 1999a; Fruchter \etal\ 1999b \\
990308 & ?     & 29.6  &  0.02 $\pm$ 1.75   &  Jaunsen \etal\ 2003 \\
990506 & 1.31  & 25.5 & -0.25 $\pm$ 1.36   & Dark; Bloom \etal\ 2003; Le Floc'h \etal\ 2003   \\
991208 & 0.71 & 24.2 &  0.34 $\pm$ 0.83   & Castro-Tirado \etal\ 2001; Fruchter \etal\ 2000b \\
991216 & 1.02 & 25.2 &  0.47 $\pm$ 0.94   & 
Vreeswijk \etal\ 1999; 2000  \\
000210 & 0.85 & 23.5 &  3.05 $\pm$ 0.76   & Dark;  Piro \etal\ 2002 \\
000301C & 2.03 & 27.9  & -1.46 $\pm$ 0.90 & Smette \etal\ 2001; Fruchter \& Vreeswijk 2001; Jensen \etal\ 2001 \\
000418 & 1.12 & 23.8 &  3.15 $\pm$ 0.90   & Bloom \etal\ 2003; Klose \etal\ 2000; Metzger \etal\ 2000  \\
000911$\dagger$ & 1.06 & 25.1 &  1.11 $\pm$ 0.63   & Price \etal\ 2002b \\
000926 & 2.04 & 24.8 &  1.40 $\pm$ 1.23   & Castro, S. \etal\ 2003; Fynbo \etal\ 2000; Rol \etal\ 2000\\
001025A & ? & 24.0 & -2.53 $\pm$ 3.04 & Dark; Pedersen \etal\ in prep.  \\
010222 & 1.48 & 25.7 &  3.74 $\pm$ 0.53   & Jha \etal\ 2001; Fruchter \etal\ 2001a; Galama \etal\ 2003 \\
010921 & 0.45 & 21.5 &  0.46 $\pm$ 1.14    & Price \etal\ 2002a; Park \etal\ 2002 \\
011211 & 2.14 & 24.8 &  1.94 $\pm$ 0.89   & Fruchter \etal\ 2001b; Holland \etal\ 2002; Jakobsson \etal\ 2003 \\
020124 & 3.20 & $>29.5$ &  1.20 $\pm$ 2.30  & Hjorth \etal\ 2003; Berger \etal\ 2002 \\
020813 & 1.25 & 24.2 & -1.40 $\pm$ 3.50         & Barth \etal\ 2003; Castro Cer\'on \etal\ A\&A in prep. \\
021004 & 2.32 & 24.3 &  0.77 $\pm$ 1.25   &  M{\o}ller \etal\ 2002; Chornock \& Filippenko 2002; Fynbo \etal\ in prep. \\
\hline
\end{tabular}

\parbox{17cm}
{
* GRB~990123 was observed by Galama \etal\ (1999) 
and Kulkarni \etal\ (1999b)
with SCUBA (850\micron)
on several occasions
between 1 to 15 days after the burst.  
Although there is the possibility of a small amount of afterglow contamination,
the flux density we list here is an weighted average of all the Galama \etal\ 
and Kulkarni \etal\ measures, together
with our rather shallow result.
For the purpose of the analysis presented here, the fact that a host flux density 
of $>2$~mJy is
ruled out is the important point.
$\dagger$ The GRB~000911 host was observed by Be03
at $2.31\pm0.91$~mJy
and also by Smith \etal\ (2001) who found $0.03\pm0.86$~mJy
about a week after the burst.
Since this is the one case of a significant discrepancy between the
Smith \etal\ result and a subsequent measurement, we have chosen to average
the two results here.\\
}
\end{table*}

\section{Discussion}
\label{sec:discussion}

The combined 
sample contains  21 host galaxies which have 850~\micron\ 
measures with better than 1.4~mJy uncertainties, indeed
most of these have errors $\lsim$ 1~mJy.  Our discussion
refers mainly to this well-observed sample.

Only three galaxies have significant positive
detections, namely the hosts of GRB~010222, GRB~000418
and GRB~000210, which are all found at $>3.5\sigma$, and
so can be regarded as confident detections.
A few others have $\sim2\sigma$ detections,
but we should be more wary of these, particularly since
there are also a couple of cases of $\sim2\sigma$
negative fluxes.
This 
latter fact could be taken as a further hint of a small underestimate
in the errors, but equally it could be a chance occurence given
the sample size and the fact that the quoted errors do not account for
crowding noise.

For reasonable dust temperatures of order 40\,K, corresponding
to restframe far-IR SEDs peaking at about 90\,$\mu$m, the luminosities
of these galaxies would be about $6 \times 10^{12}$\,L$_\odot$. If all
this energy was provided by star formation, then a star formation
rate upward of 1000\,M$_\odot$\,yr$^{-1}$ would be required
(eg. Blain, Barnard and Chapman 2003).

Two hosts previously identified as possibly highly
star forming are GRB~980703 and GRB~000911, on the
basis of radio flux (Berger \etal\ 2001b) and submm (Be03) respectively.
Our compilation suggests that in both instances, these
initial results were overestimates -- in the former case, plausibly
due to the scatter in the FIR-Radio correlation.
In the case of GRB~000911, the conclusion proceeds
from the low flux seen in the afterglow observations
of Smith \etal\ (2001).
As a general point, we note that when working at the limit
of detection,
since there are many more low luminosity galaxies
than high luminosity, 
the fluxes of
the brightest galaxies in any
sample are more likely to be overestimates than underestimates.
This effect is akin to to Eddington-Malmquist bias, and tells us that
we should not be surprised if the fluxes of the brighter galaxies
are frequently found to be lower on remeasurement.

\subsection{Properties of the sample}

For these 21 hosts, the error-weighted mean 850\micron\ 
flux density is $0.93\pm0.18$~mJy.
This can be used to test the null hypothesis of
zero flux for the sample, and, as expected, rejects
it with confidence. 
However, this number is not a fair estimate of the true mean of the sample
since we have not accounted for the intrinsic dispersion
in GRB host galaxy luminosities (about which we do not have
prior knowledge), and also because the brighter galaxies were 
in some cases (notably GRB~010222)
observed for longer and hence have smaller error
bars for that reason.  
The unweighted mean is $0.58\pm0.36$~mJy, and is a fairer
estimate of the true mean.
This is higher than the average 850\micron\ flux density found for
samples of Lyman-break galaxies, which range between
0 and $\sim$0.4~mJy, depending on the exact sample selection
(Chapman \etal\ 2000; Peacock \etal\ 2000; Webb \etal\ 2003).

What distribution do we expect to see?
As a starting point, we assume
that both submm luminosity and GRB rate are perfectly correlated with
star formation rate.
In that case, if submm flux and GRB detection completeness
were also perfectly redshift 
independent, then the predicted distribution would simply
be the submm-luminosity-weighted luminosity function of all
galaxies.

Ramirez-Ruiz, Trentham \& Blain (2002) 
use the models of Blain \etal\ (1999a,b)
for the evolution of the submm galaxy population to predict the the 
850\micron\ flux density distribution of GRB hosts.
Uncertainties in the model are almost entirely due to the uncertain
link between the GRB rate and high-mass star formation heating dust.
Subject to the, probably low, rate of contamination in the submm
galaxy population from AGN heating, and to possible evolution of the
initial mass function (IMF), the flux density distribution of submm
galaxies is reasonably well constrained throughout the interval
1--15\,mJy (Blain \etal\ 2002; Borys \etal\ 2003). Although it
predates the substantially complete observed redshift
distribution of submm galaxies from
Chapman \etal\ (2003a), the assumed model is in good agreement with
these results.
It is possible that the galaxies are 
systematically cooler and less luminous (eg. Efstathiou \& Rowan-Robinson 2003),
however, 
if that is the case, then the redshift distribution is 
only consistent with significant evolution of the form 
of the far-infrared--radio correlation with redshift;

Figure 1 shows the predictions of this model as filled points.
To provide a fair comparison, the model output was convolved with
with the distribution of observational
errors, 
and includes
a correction to allow for the
possibility of the chance appearance of
an unrelated submm source in the SCUBA beam.
The latter correction amounts to roughly 0.7 of an
object contaminating the 2--4~mJy bin 
and is based on the source counts summarised in Blain \etal\ (2002).
We have chosen
to put error bars on the model, reflecting the counting statistics
of the proportion expected in each bin.
The histogram represents the submm observations.

\begin{figure}
\centerline{\psfig{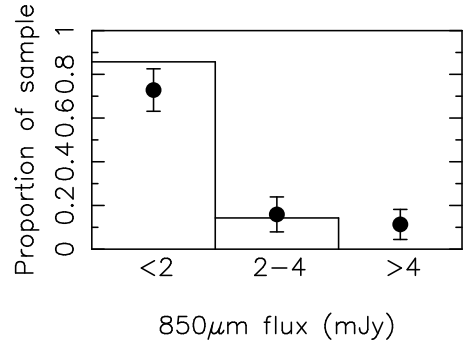}}
\caption{
Comparison between observations (histogram) 
for the 21 hosts with $<1.4$~mJy flux uncertainties,
and predictions  (shown as filled points).
The results are binned into just 3 (different sized) bins,
essentially representing no confident detection, 
confident but faint detection, and ``bright'' detection
respectively
The predictions are based on those from 
Ramirez-Ruiz, Trentham \& Blain (2002)
but also account for the observational error distribution
and the crowding noise (ie. essentially the rate of false positives).
Errors bars are placed on the predictions simply reflecting the
counting statistics rather than any uncertainty in the model
parameters.
In that sense, they are a lower limit to the true errors.
}

\label{figpred}
\end{figure}

It is apparent that there is a dearth of hosts with
$>4$ mJy 850 \micron\ flux densities, which is becoming statistically
significant (formally, at the 1.6$\sigma$ level).  
In fact, the significance is greater
if we consider the remaining 5 galaxies with larger
error bars, namely 
GRB~970828, GRB~990308, GRB~001025A, 
GRB~020124 and GRB~020813
(and indeed one might also include GRB~980519 from Smith \etal\
1999), since 
at $1\sigma$ these are all inconsistent with
being $>4$~mJy sources  as well.

Furthermore, we note that if the estimated flux 
uncertainties are somewhat optimistic
(as suggested in section 3),
it tends to make the discrepancy between theory and
observations rather worse.
This seems counterintuitive, but is due to the
fact that larger observational errors should result
in more low luminosity systems appearing in
the brighter bins by chance.

At this stage there are many possible explanations for
the discrepancy (beyond the relatively small sample size).
First and foremost, we must still worry 
that amongst the GRBs with ``very dark'' afterglows,
for which good radio or X-ray positions are not available,
lurk a small number of highly submm luminous hosts. 
Another potential
selection effect is that GRBs are being
preferentially picked up at lower redshifts, both in terms of initial 
detection, but more significantly the detection of
the afterglows, as discussed in section 1.3.  
This would be a less direct
selection against the higher redshift, dusty
systems, although we note that of the seven systems
with $z>1.5$, {\em none} have significant 850\micron\ fluxes.

Other possible physical explanations are: 
(a) 
that GRBs are preferentially found in low metallicity
-- smaller and less dusty -- systems.
This has been predicted as a consequence of the  
single-star collapsar
model (Heger \etal\ 2003), 
and is consistent with 
the apparent enhanced brightness of GRBs 
in the outer parts of their parent galaxies
(Ramirez-Ruiz, Lazzati \& Blain 2002), and
the relatively low metallicity 
inferred for many GRB hosts 
(eg. 
Fynbo \etal\ 2003).
The latter study notes that a
a large proportion of GRB hosts are Ly$\alpha$
emitters (five out of five so far studied), which is
significantly greater than the proportion of Lyman-break
galaxies, although perhaps surprisingly, many of the
SCUBA galaxies for which Chapman \etal\ have recently
acquired redshifts, are also strong Ly$\alpha$ emitters; 
(b) 
that the higher luminosity
submm galaxies are more contaminated by AGN, or the submm
luminosity function more contaminated by cooler, less luminous galaxies,
than 
has been generally
thought;
(c) 
that whilst both GRBs and submm flux trace star-formation
rate, the two are not perfectly correlated due to different
phase lags with respect to the true star formation rate.
For instance, the simulation of Bekki and Shioya (2001)
of a merger induced starburst, shows significant fluctuations in the star-formation
rate on timescales much less than 100 Myr (their fig 6).
Given that both GRB rate and submm flux will not follow
rapid fluctuations instantaneously, some decorrelation
may occur.
(d) 
that variations in the stellar IMF
are having a different effect on GRB rate
compared to submm flux; or
(e) 
that GRBs
for some reason are occuring preferentially in 
galaxies with high dust temperatures, so more of the
bolometric luminosity appears at shorter wavelengths and
is being missed at 850\micron.

\subsection{Properties of the submm bright GRB hosts}
\label{subsec:submmbrthosts}

If submm flux traces star formation, and the selection
of GRB hosts is unbiased, then the submm-bright GRB
hosts as a whole should be similar to the general
population of submm bright galaxies.
Of course, now we are dealing with a very small sample, but it
is interesting to examine the properties of the three
hosts with 850\micron\ flux densities above 2~mJy.

Gorosabel \etal\ (2003a)
have constructed a UV to IR SED for the host to
GRB~000210, which is well fit by a starburst SED
with a relatively 
modest star formation rate of 2~\Msol~\per{yr}
and
negligible extinction.
The redshift is $z=0.84$ and the luminosity $L\approx0.5L_{\star}$.

The UV to IR SED of the GRB~000418 host (Gorosabel \etal\ 2003b) is 
also well fit by a starburst template, in this case with 
a moderate amount of extinction $A_V\sim0.4$, and a star
formation rate up to 60~\Msol~\per{yr}.
A similar result is obtained in the spectroscopic study
of Bloom \etal\ (2003), who also find no evidence of
any AGN contamination.
This star formation rate is still an order of magnitude less than that derived
from the submm, but that is not unreasonable for heavily
obscured galaxies.
The redshift is $z=1.12$ and
the luminosity is $L\approx L_{\star}$. 
It is compact
with effective radius 
$r_E\approx 1$ kpc,
and the afterglow was located close to the optical centroid
(Fruchter \& Metzger 2001; Bloom \etal\ 2003).
The afterglow itself has produced extinction
estimates ranging from  $A_V\sim0.4$ (Berger \etal\ 2001a)
to  $A_V\sim1$
(Klose \etal\ 2000).

The host of GRB~010222 is fainter still in the optical
at $R\approx25.3$
(Fruchter \etal\ 2001a), but it is also moderately blue, with 
$I-K\approx2.1$ and an intrinsic luminosity $L\sim0.1L_{\star}$ (Frail \etal\ 2002). 
The redshift is $z=1.477$ and again {\it HST} images reveal the afterglow
to have been located on top of the peak of the optical emission.

In all these cases if the submm is really indicating copious amounts of
hidden star formation activity, 
it would appear that the optically visible 
part of the host is dominated by one or more
relatively unextinguished regions.
This is plausible -- it could be concentrated in a shell or plume of material, 
or possibly a number of ``windows'' through an
obscuring shroud (eg. Bekki \etal\ 1999).
However, if this is the case, it also seems that the optically 
bright regions must be spatially proximate to the bright far-IR
emission if we are to understand why the afterglows (as with most
detected GRBs) tend to be found close to the optical centroids (or hotspots)
of the hosts.

So, how do the
properties of these galaxies compare with the 
the optical/IR counterparts of the submm bright galaxies 
selected in blank-field SCUBA surveys?
In fact the bright SCUBA galaxies display
quite a wide variety of characteristics.  
They have been split into three classes
(Smail \etal\ 2002): class 0 are relatively bright in
optical as well as submm -- much of the star formation is
unextinguished; class 1 are extremely red objects (EROs) with
$I-K$ typically greater than 5; finally class 2
are extremely faint in optical and IR.
This classification system was conceived based on 
galaxies which 
are generally brighter than 5~mJy at 850\micron.
Although no GRB host to-date is that bright, it appears
the the submm brightest of the GRB hosts do not 
fit neatly into this scheme, being intermediate in 
optical/IR luminosity between classes 0 and 2, but not very
red EROs like class 1.
Interestingly, the best candidate for an extremely red GRB host galaxy, 
that of GRB~980326
which has $R-K\gsim6$ (although the $K$ detection at $K=22.9$
is $<3\sigma$; Chary, Becklin \& Armus 2002), 
has a low submm luminosity.
This compares to an average 850\micron\ flux density of $\approx 1.6$~mJy
recently found by Wehner, Barger \& Kneib (2002) for a
$K$-selected sample of EROs.

Furthermore, as discussed in section 1.2, the redshift
distribution of bright SCUBA sources shows a peak
in the redshift range 2--3, whereas these 3 submm bright
GRB hosts are nearer $z\sim1$.

We conclude that the properties of the few submm bright GRB
hosts found to-date 
are not very representative of the submm-bright
galaxies as a whole: they do not fit easily within
any of the 3 classes.  
Be03 reached the same conclusion based on a colour--redshift
plot, in which the 
submm-bright GRB hosts lie bluer and at lower redshift
than the submm selected galaxies as a whole.
This could be because the GRB hosts are somewhat
fainter at 850\micron, since nearly
all well studied submm samples are $\gsim5$~mJy (although 
Frayer \etal\ 2003 report a detailed study of a single 
gravitationally lensed source 
at $z=2.51$, which has a low submm flux density of 
$S_{850}<2$~mJy after correction for lensing).
As discussed above, we also
expect GRB selection to provide some bias toward picking up 
lower redshift galaxies, and against
very dusty galaxies.
The alternative, of course, is that either the submm selection has
some problems, such as a surprising rate of AGN contamination, or
surprising dust properties, or GRB progenitors are more likely
to arise in less dusty systems.

\section{Conclusions}
\label{sec:conclusions}

We have shown that whilst a small number of GRB hosts
are bright submm galaxies, indicating high star formation
rates, the proportion, particularly at flux densities $>4$~mJy, 
is fewer than predicted  if both GRBs and submm flux
trace star formation in an unbiased way.  Furthermore, and similarly
puzzling, the three submm-brightest GRB hosts are not very
typical in terms of their optical properties and redshift
distribution of the galaxies selected in blind submm surveys.

Of course, the
current sample size is still rather small, 
and very likely suffers some
selection biases.
In particular, GRBs in very dusty hosts and/or at higher
redshifts are more likely to be missing from the observed
afterglow samples.  A small number of these amongst the
``dark'' bursts (those without optical counterparts to
deep limits) could remove the discrepancy -- although
the  afterglows known to be genuinely very
faint is only a small fraction of those for which optical
emission has not been detected.  Furthermore, 
a fair proportion of the darker bursts
seem to be dark for reasons
other than their residing in high redshift, dusty ULIRGs,
so it's not obvious that this selection effect provides the 
full explanation. 

Future studies may largely
overcome these 
selection problems by much earlier and more uniform afterglow
searches with {\it SWIFT} and various robotic telescopes. 
The positions from the {\it SWIFT}/XRT 
may be good enough to identify
ULIRG hosts, even when no optical afterglow is detected.
{\it Spitzer} will also be a powerful tool, particularly the
for the lower redshift hosts $z\lsim1$, where the large majority
of the bolometric luminosity appears in the far-IR. 

If it turns out that the discrepancy with the
model predictions persists
when more
complete samples are studied, it will certainly 
be telling us something interesting about the astrophysics
of GRBs and/or the relationship between FIR emission and
star formation.
On the other hand, if the disagreement goes away,
the use of GRBs as practical 
star formation indicators will be strongly bolstered.

\section*{Acknowledgements}

We thank the referee, Steve Serjeant, for insightful comments,
Jochen Greiner for keeping his invaluable website up to date 
(http://www.mpe.mpg.de/$\sim$jcg/grbgen.html) and Paul Vreeswijk, Jens Hjorth,
Javier Gorosabel and Johan Fynbo for useful discussions.
We also thank all the 
observers and staff at the JCMT who carried out flexible-scheduled 
observations on our behalf.
The JCMT is operated by the Joint Astronomy 
Centre on behalf of the Particle Physics and Astronomy
Research Council of the UK, the Netherlands Organization
for Scientific Research, and the National Research Council
of Canada.

\end{document}